\documentclass[12pt, draftclsnofoot, onecolumn]{IEEEtran}
\usepackage[T1]{fontenc}
\usepackage[pdftex]{graphicx}
\usepackage{epstopdf}
\usepackage{transparent}
\usepackage{eso-pic}

\usepackage{amsfonts, amscd, mathrsfs, amsopn, bbm, bbold, multicol, relsize}
\usepackage{mathrsfs}
\usepackage[psamsfonts]{amssymb}
\usepackage{mathbbol}
\usepackage[mathscr]{eucal}
\usepackage[cmex10]{amsmath}
\usepackage[amsmath,thmmarks]{ntheorem}
\makeatletter
\renewtheoremstyle{plain}
{\item{\theorem@headerfont ##1\ ##2\theorem@separator}~}
{\item{\theorem@headerfont ##1\ ##2\ (##3)\theorem@separator}~}
\makeatother

{\theoremheaderfont{\upshape\bfseries}
	\theorembodyfont{\normalfont\slshape}
	\newtheorem{definition}{Definition}}

\usepackage{mathtools}
\usepackage{cite}
\usepackage{tikz}
\usepackage{graphicx}

\theoremstyle{plain}
\theoremheaderfont{\itshape\bfseries}
\theorembodyfont{\normalfont\itshape}
\theoremseparator{:}
\theoremsymbol{}
\newtheorem{theorem}{Theorem}
\newtheorem{lemma}{Lemma}

\newtheorem{remark}{Remark}

\theoremstyle{nonumberplain}

\theoremstyle{nonumberplain}
\theoremheaderfont{\itshape}
\theorembodyfont{\normalfont}
\theoremseparator{:}
\theoremsymbol{}

\theoremstyle{plain}
\theoremheaderfont{\itshape\bfseries}
\theorembodyfont{\normalfont}
\theoremseparator{:}

\theoremstyle{nonumberplain}
\theoremsymbol{\rule{1ex}{1ex}}

\newlength\fheight
\newlength\fwidth
\DeclarePairedDelimiter\abs{\lvert}{\rvert}

\newcommand\restr[2]{{
  \left.\kern-\nulldelimiterspace 
  #1 
  \right|_{#2} 
  }}

\newcommand{\Ac}{{\cal A}}
\newcommand{\Bc}{{\cal B}}

\newcommand{\Dc}{{\cal D}}

\newcommand{\Mc}{{\cal M}}

\newcommand{\Uc}{{\cal U}}

\newcommand{\Vc}{{\cal V}}

\usepackage{tikz}

\usepackage[utf8]{inputenc}
\usepackage{pgfplots} 
\usepackage{pgfgantt}
\usepackage{pdflscape}
\pgfplotsset{compat=newest} 
\pgfplotsset{plot coordinates/math parser=false}
\usepackage{relsize}
\usepackage{color}
\usepackage{changes}
\usepackage{graphicx,color,epsfig,rotating,subfigure}
\usepackage[cmintegrals]{newtxmath}
\begin{document}
	
	\title{On D2D Caching with Uncoded Cache Placement}
	
	\author{
		\c{C}a\u{g}kan~Yapar,~\IEEEmembership{Student~Member,~IEEE,} 
		Kai~Wan,~\IEEEmembership{Member,~IEEE,} \\
		Rafael F. Schaefer,~\IEEEmembership{Senior~Member,~IEEE,}
		and~Giuseppe Caire,~\IEEEmembership{Fellow,~IEEE}
		
		\thanks{
			\c{C}a\u{g}kan~Yapar, Kai Wan and Giuseppe Caire are with the Communications and Information Theory Chair, Technische Universit\"at Berlin, Germany
			(e-mail: cagkan.yapar@tu-berlin.de; kai.wan@tu-berlin.de; caire@tu-berlin.de). The work of \c{C}. Yapar, K.~Wan and G.~Caire was funded by the European Research Council  under the ERC Advanced Grant N. 789190, CARENET.}
		
		\thanks{Rafael F. Schaefer was with the Information Theory and Applications Chair, Technische Universit\"at Berlin, Germany. Currently, he is with the Chair of Communications Engineering and Security, University of Siegen, Germany (email: rafael.schaefer@uni-siegen.de). The work of R. F. Schaefer was funded by the German Ministry of Education and Research (BMBF) within the national initiative for “Post Shannon Communication (NewCom)” under Grant 16KIS1004.
		}
 		
	}

	\IEEEoverridecommandlockouts
	\maketitle
	
\begin{abstract}
	We consider a cache-aided wireless device-to-device (D2D) network under the constraint of \emph{one-shot delivery}, where the placement phase is orchestrated by a central server. We assume that the devices' caches are filled with uncoded data, and the whole file database at the server is made available in the collection of caches. Following this phase, the files requested by the users are serviced by inter-device multicast communication. For such a system setting, we provide the exact characterization of load-memory trade-off, by deriving both the minimum average and the minimum peak sum-loads of links between devices, for a given individual memory size at disposal of each user.
\end{abstract}

\section{Introduction}
The killer application of wireless networks has evolved from real-time voice communication to on-demand multimedia content delivery (e.g., video), which requires a nearly $100$-fold increase in the per-user throughput, from tens of kb/s to  $1$ Mb/s.  Luckily, the pre-availability of such content allows for leveraging storage opportunities at users in a proactive manner, thereby reducing the amount of necessary data transmission during periods of high network utilization.

A caching scheme is composed of two phases. The \emph{placement phase} refers to the operation during low network utilization, when users are not requesting any content. During this phase, the cache memories of users are filled by a central server proactively. When each user directly stores a subset of bits, the placement phase is uncoded.
The transmission stage when users request their desired content is termed \emph{delivery phase}. By utilizing the content stored in their caches during the placement phase, users aim to reconstruct their desired content from the signals they receive. The sources of such signals may differ depending on the context and network topology. In this work, we focus on the device-to-device (D2D) caching scenario, in which the signals available during the delivery phase are generated merely by the users themselves, whereas the central server remains inactive. 

A coded caching strategy was proposed  by Maddah-Ali and Niesen (MAN) \cite{maddah2014fundamental}. Their model consists of users with caches and of a server which is in charge of the distribution of content to users through an error-free shared-link, during both the placement and delivery phases. This seminal work showed that a \emph{global caching gain} is possible by utilizing   multicasting linear combinations during the delivery phase. 
By observing that some MAN linear combinations are redundant, the authors~\cite{yu2018exact} proposed an improved scheme, which is optimal under the constraint of uncoded cache placement. It was proved in~\cite{yu2017characterizing} that the uncoded caching scheme is optimal generally within a factor of $2$, e.g. even when more involved (coded) cache placement schemes are allowed.

D2D caching problem was originally considered in \cite{ji2013fundamental,ji2016fundamental,ibrahim2018device}, where  users are allowed to  communicate with each other. By extending  the caching scheme in \cite{maddah2014fundamental} to the D2D scenario, we can also have the  global caching gain. It was proved in \cite{ji2013fundamental,ji2016fundamental} that the proposed D2D caching scheme is order optimal within a constant when the memory size is not small.

Particularly, the D2D caching setting with uncoded placement considered in this work is closely related to the  distributed computing  problem originally proposed  \cite{li2018fundamental} and  distributed data-shuffling problem  \cite{wan2018fundamental}. The coded distributed computing setting can be interpreted as a symmetric D2D caching setting with multiple requests, whereas the coded data shuffling problem can be viewed as a D2D caching problem with additional constraints on the placement.



\paragraph*{\textbf{Contributions}}
Based on the D2D achievable caching scheme in~\cite{ji2016fundamental}, with $K$ the number of users and $N$ the number of files, for $N\geq K$ and the shared-link caching scheme in~\cite{yu2018exact} for $N<K$,
we propose a novel achievable scheme for D2D caching problem, which is shown to be order optimal within a factor of $2$ under the constraint of uncoded placement, in terms of the average transmitted load for uniform probability of file requests and the worst-case transmitted load among all possible demands.

For each user, if any bit of its demanded file not already in its cache can be recovered from its cache content and a transmitted packet of a single other user, we say that the delivery phase is \emph{one-shot}.
Under the constraint of uncoded placement and one-shot delivery, we can divide the D2D caching problem into $K$ shared-link models. Under the above constraints, we then use the index coding acyclic converse bound in~\cite[Corollary 1]{onthecapacityindex}  to lower bound the total load transmitted in the $K$ shared-link models. By leveraging the connection among the $K$ shared-link models, we propose a novel way to use the index coding acyclic converse bound compared to the method used for single shared-link model in~\cite{wan2016optimality,wan2016caching,yu2018exact}. With this converse bound, we prove that the proposed achievable scheme is exactly optimal under the constraint of uncoded placement and one-shot delivery, in terms of the average transmitted load and the worst-case transmitted load among all possible demands.

In the longer version of this work \cite{arXivVersion}, we also consider random user inactivity where the identities of inactive users are unknown. The one-shot delivery property allows for an extension of the proposed scheme that provides robustness against outage that user inactivity may lead to.

\section{Problem Setting}\label{sec:sys}

\subsection{Notations}
\label{sec:model:notation}
$|\cdot|$ is used to represent the cardinality of a set or the length of a file in bits;
we let 
$\mathcal{A\setminus B}:=\left\{ x\in\Ac|x\notin\mathcal{B}\right\}$,
$[a:b:c]:=\{a,a+b,a+2b,...,c\}$,
$[a:c] = [a:1:c]$ and $[n]=[1:n]$;
the bit-wise XOR operation between binary vectors is indicated by $\oplus$;
for two integers $x$ and $y$, we let $\binom{x}{y}=0$ if $x<y$ or $x\leq0$.

\subsection{D2D Caching Problem Setting}
\label{sub:problem setting}
We consider a D2D network composed of $K$ users, which are able to receive all the other users' transmissions. 
Users make requests from a   database of $N$ files $\boldsymbol{W} = (W_1,\dots, W_N)$, each with a length of $F$ bits. Every user has a memory of $MF$ bits, where $ M \in [N/K,N)$. 

The system operation can be divided into  the \emph{placement} and \emph{delivery} phases.
During the placement phase users have access to a central server.  In this work, we only consider the caching problem with uncoded cache placement, 
where each user $k$ directly stores  $MF$ bits of $N$ files in its memory. For the sake of simplicity, we do not repeat this constraint in the rest of paper.
Since the placement is uncoded, we can divide each file into subfiles, $W_{q}=\{W_{q,\Vc}:\Vc\subseteq [K]\}$, where $W_{q,\Vc}$ represents the set of bits exclusively cached by users in $\Vc$.
We denote the indices of the stored bits at user $k$ by $\mathcal{M}_k$.  For convenience, we denote the cache placement of the whole system by $\boldsymbol{\mathcal{M}} := (\mathcal{M}_1,\dots,\mathcal{M}_K)$. 
During the delivery phase, each user demands one file. We define \textit{demand} vector $\boldsymbol{d}:=\left(d_1,\dots,d_K\right)$, with $d_k \in [N]$ denoting user $k$'s requested file index. The set of all possible demands is denoted by $\mathcal{D}$, so that $\mathcal{D}=[N]^K$.
Given the demand information, each user $k$ generates a codeword $X_k$ of length $R_k F$ bits and broadcasts it to other users, where $R_k$ indicates the load of user $k$. For a given subset of users $\mathcal{S} \subseteq [K]$, we let $X_{\mathcal{S}}$ denote the ensemble of codewords broadcasted by these users. From $\mathcal{M}_k$ and   $X_{[K]\backslash k}$, each user $k$  recovers its desired file.

In this work we concentrate on the special case of \emph{one-shot delivery}, which we formally define in the following. 
\begin{definition}[One-shot delivery]
	If each user $k \in [K]$ can decode any bit of its requested file not already in its own cache 
	from its cache and the transmission of a single other user, we say that the delivery phase is {\em one-shot}. Mathematically, we indicate by $W^{k,i}_{d_k}$ the block of bits needed by user $k$ and recovered from the transmission of user i, i.e., 
	$ H(W^{k,i}_{d_k} | X_i, \Mc_k) = 0,$
	indicating that  $W^{k,i}_{d_k}$ is a deterministic function of $X_i$ and $\Mc_k$. 
	Then, a one-shot scheme implies that 
	$ (W_{d_k} \setminus \Mc_k) \subseteq \bigcup_{i \in [K]\setminus \{k\}} W^{k,i}_{d_k}.$
	In addition, we also define $W^{k,i}_{d_k,\Vc}$ as the block of bits needed by user $k$ and recovered from the transmission of user $i$, which are exclusively cached by users in $\Vc$. Hence, we have for each user $k\in[K]$
	$ \bigcup_{\Vc\subseteq ([K]\setminus \{k\}): i\in \Vc} W^{k,i}_{d_k,\Vc}=W^{k,i}_{d_k}, \forall i\in [K]\setminus \{i\}.$
\end{definition}


Letting $R = \sum_{k=1}^{K} R_k$, we say that a communication load $R$ is \textit{achievable} for a demand $\boldsymbol{d}$ and placement $\boldsymbol{\mathcal{M}}$, with $\abs{\mathcal{M}_k}=M,\, \forall k \in [K]$, if  and only if there exists an ensemble of codewords $X_{[K]}$ of size $RF$ such that each user $k$ can reconstruct its requested file $W_{d_k}$. We let $R^*(\boldsymbol{d},\boldsymbol{\mathcal{M}})$ indicate the minimum achievable load given $\boldsymbol{d}$ and $\boldsymbol{\mathcal{M}}$.
We also define $R^*_{\textup{o}}(\boldsymbol{d},\boldsymbol{\mathcal{M}})$ as the minimum  achievable load given $\boldsymbol{d}$ and $\boldsymbol{\mathcal{M}}$ under the constraint of one-shot delivery. We consider uniform demand distribution and aim to minimize the average and worst-case loads, $R^*_{\textup{ave}}=\min_{\substack{\boldsymbol{\mathcal{M}}}} \mathbb{E}_{\boldsymbol{d}}[ R^*(\boldsymbol{d},\boldsymbol{\mathcal{M}})]$ and $R^*_{\textup{worst}}=\min_{\boldsymbol{\mathcal{M}}} \max_{\boldsymbol{d}} R^*(\boldsymbol{d},\boldsymbol{\mathcal{M}})$.
Similarly, we define $R^*_{\textup{ave, o}}$ and  $R^*_{\textup{worst, o}}$ as the minimum average and worst-case loads under the constraint of one-shot delivery, respectively.

Further, for a demand $\boldsymbol{d}$, we let $N_{\textup{e}}(\boldsymbol{d})$ denote the number of distinct indices in $\boldsymbol{d}$. In addition,
we let $\boldsymbol{d}_{\backslash\{k\}}$ and $N_{\textup{e}}(\boldsymbol{d}_{\backslash\{k\}})$ stand for the demand vector of users $[K]\backslash\{k\}$ and the number of distinct files requested by all users but user $k$, respectively. 
As in \cite{tian2016symmetry,yu2018exact}, we group the demand vectors in $\mathcal{D}$ according to the frequency of common entries that they have. Towards this end, for a demand $\boldsymbol{d}$, we stack in a vector of length $N$ the number of appearances of each request in descending order, and denote it by $\boldsymbol{s}(\boldsymbol{d})$. 
We refer to this vector as \textit{composition} of $\boldsymbol{d}$. Clearly, $\sum_{n=1}^{N}s_n(\boldsymbol{d})=K$.  By $\mathscr{S}$ we denote the set of all possible compositions. We denote the set of demand vectors with the same composition $\boldsymbol{s} \in \mathscr{S}$  by $\mathcal{D}_{\boldsymbol{s}}$. We refer to these subsets as \textit{type}s. Obviously, they are disjoint and $\bigcup\limits_{\boldsymbol{s}\in \mathscr{S}} \mathcal{D}_{\boldsymbol{s}} = \mathcal{D}$. 

\section{Main Results}\label{sec:RMtrade-off}

In the following theorem, we characterize the exact memory-average load trade-off under the constraint of one-shot delivery. The achievable scheme is introduced in Section~\ref{sec:achiev} and the converse bound is proved in Section~\ref{sec:Converse}.
\begin{theorem}[Average load]\label{teo}
	For a D2D caching scenario with a database of $N$ files and $K$ users each with a cache of size $M$,  the following average load under the constraint of uncoded placement and one-shot delivery with uniform demand distribution, is optimal
	\begin{equation}
	R^*_{\textup{ave, o}}=\mathbb{E}_{\boldsymbol{d}}\left[ \frac{\binom{K-1}{t}-\frac{1}{K}\sum_{k=1}^K\binom{K-1-N_{\textup{e}}(\boldsymbol{d}_{\backslash\{k\}})}{t}}{\binom{K-1}{t-1}}\right]\label{eq:averageworstcase}
	\end{equation}
	with $t=\frac{KM}{N} \in [K]$, where $\boldsymbol{d}$  is uniformly distributed over $\mathcal{D}=\{1,...,N\}^K$. Additionally, $R^*_{\textup{ave, o}}$ corresponds to the lower convex envelope of its values at $t\in [K]$, when $t\notin [K]$.
\end{theorem}

We can  extend the above results to worst-case  load in the following theorem, whose proof can be found in \cite{arXivVersion}.
\begin{theorem}[Worst-case load]
	\label{corr}
	For a D2D caching scenario with a database of $N$ files and $K$ users each with a cache of size $M$ ,  the following peak load $R^*_{\textup{worst, o}}$ under the constraint of uncoded placement and one-shot delivery, is optimal
	\begin{align}
	R^*_{\textup{worst, o}}= \begin{cases} \frac{\binom{K-1}{t}}{\binom{K-1}{t-1}} & K \leq N  \\  
	\frac{\binom{K-1}{t}-\frac{2N-K}{K}\binom{K-N}{t}-\frac{2(K-N)}{K}\binom{K-1-N}{t}}{\binom{K-1}{t-1}} & \textup{otherwise}\\
	\frac{\binom{K-1}{t}-\binom{K-1-N}{t}}{\binom{K-1}{t-1}} & K \geq 2N
	\end{cases}\label{eq:worstCaseload}
	\end{align}
	with $t = \frac{KM}{N} \in [K]$. Additionally, $R^*_{\textup{worst, o}}$ corresponds to the lower convex envelope of its values at $t\in [K]$, when $t\notin [K]$.
\end{theorem}

\begin{remark}
	\label{rem:symmetry in file-splitting}
	As we will present in Section \ref{sec:achiev} and discuss in Remark \ref{rem:Ksharedlink}, our achievable scheme is in fact composed of $K$ shared-link sub-systems, where each $i^{\textup{th}}$ sub-system includes $N_{\textup{e}}(\boldsymbol{d}) = N_{\textup{e}}(\boldsymbol{d}_{\backslash\{i\}})$ demanded files. The scheme is symmetric in the file-splitting step in the delivery phase. It is interesting to observe that, even if the system is asymmetric in the sense that  each $K$ shared-link sub-system may not have the same $N_{\textup{e}}(\boldsymbol{d}_{\backslash\{i\}})$, the symmetric file-splitting is nevertheless optimal.
\end{remark}
By comparing the achievable load by our proposed scheme and the minimum achievable load for shared-link model, we obtain the following order optimality result (see the longer version \cite{arXivVersion} of this paper for the proof).
\begin{theorem}[Order optimality]\label{thm:order optimality}
	For a D2D caching scenario with a database of $N$ files and $K$ users each with a cache size of $M$, the proposed achievable average and worst-case transmitted loads in~\eqref{eq:averageworstcase} and~\eqref{eq:worstCaseload}, is order optimal within a factor of $2$.
\end{theorem}

\section{A Novel Achievable D2D Coded Caching Scheme}\label{sec:achiev}
In this section, we present a caching scheme that achieves the loads stated in Theorem \ref{teo} and Theorem \ref{corr}. To this end, we show that for any demand vector $\boldsymbol{d}$ the proposed scheme achieves the load
\begin{align}\label{eq:single}
R^*(\boldsymbol{d},\boldsymbol{\boldsymbol{\mathcal{M}}_\textup{MAN}})=\frac{\binom{K-1}{t}-\frac{1}{K}\sum_{i=1}^{K}\binom{K-1-N_{\textup{e}}(\boldsymbol{d}_{\backslash\{i\}})}{t}}{\binom{K-1}{t-1}},
\end{align}
where $\boldsymbol{\mathcal{M}}_\textup{MAN}$ refers to the symmetric placement which was originally presented in \cite{maddah2014fundamental}. This immediately proves the achievability of the average and worst case loads given in Theorem \ref{teo} and Theorem \ref{corr}, respectively. In Subsection \ref{sub:achiev}, we will present our achievable scheme and provide a simple example, illustrating how the idea of exploiting common demands \cite{yu2018exact} is incorporated in the D2D setting. In Remark~\ref{rem:Ksharedlink}, we will discuss our approach of decomposing the D2D model into $K$ shared-link models.

\subsection{Achievability of $R^*(\boldsymbol{d},\boldsymbol{\boldsymbol{\mathcal{M}}_\textup{MAN}})$}\label{sub:achiev}
In the following, we present the proposed caching scheme for integer values of $t \in [K]$. For non-integer values of $t$, resource sharing schemes \cite{maddah2014fundamental,maddah2015decentralized,ji2016fundamental} can be used  to achieve the lower convex envelope of the achievable points.
\subsubsection{Placement phase}
Our placement phase is based on the MAN placement \cite{maddah2014fundamental}, where each file $W_q$ is divided into $\binom{K}{t}$ disjoint sub-files denoted by $W_{q, \Vc}$ where $\Vc \subseteq [K]$ and $|\Vc| = t$. During the placement phase, each user $k$ caches all bits in each sub-file $W_{q, \Vc}$ if $k \in \Vc$. As there are $\binom{K-1}{t-1}$ sub-files for each file where $k \in \Vc$ and each sub-file is composed of $F/\binom{K}{t}$ bits, each user caches $NFt/K = MF$ bits.
\subsubsection{Delivery phase}
The delivery phase starts with the \emph{file-splitting} step: Each sub-file is divided into $t$ equal length disjoint sub-pieces of $F/ t \binom{K}{t}$ bits which are denoted by $W_{q, \Vc, i}$, where $i\in \Vc$.
Subsequently, each user $i$ selects any subset of $N_{\textup{e}}(\boldsymbol{d}_{\backslash\{i\}})$ users from $[K]\backslash\{i\}$, denoted by $\mathcal{U}^i=\{u_1^i,...,u^i_{N_{\textup{e}}(\boldsymbol{d}_{\backslash\{i\}})}\}$, which request $N_{\textup{e}}(\boldsymbol{d}_{\backslash\{i\}})$ distinct files. Extending the nomenclature in \cite{yu2018exact}, we refer to these users as \textit{leading demanders of user $i$}. 

Let us now fix a user $i$ and consider an arbitrary subset $\mathcal{A}^i\subseteq [K]\backslash\{i\}$ of $t$ users. Each user $k\in \mathcal{A}^i$ needs the sub-piece $W_{d_k, \{\mathcal{A}^i \cup \{i\}\}\backslash\{k\}, i}$, which is cached by all the other users in $\mathcal{A}^i$ and the user $i$. Precisely, all users in a set $\mathcal{A}^i$ wants to exchange these sub-pieces $W_{d_k, \{\mathcal{A}^i \cup \{i\}\}\backslash\{k\}, i}$ from the transmissions of user $i$. By letting user $i$ broadcast 
\begin{equation}\label{eq:Broadcasts}
Y_{\mathcal{A}^i}^i:=\underset{k\in \mathcal{A}^i}{{\bigoplus}}W_{d_k, \{\mathcal{A}^i \cup \{i\}\}\backslash\{k\}, i},
\end{equation} 
this sub-piece exchanging can be accomplished, as each user $k \in \Ac^i$ has all the sub-pieces on the RHS of \eqref{eq:Broadcasts}, except for $W_{d_k, \{\mathcal{A}^i \cup \{i\}\}\backslash\{k\}, i}$.

We let each user $i$ broadcast the binary sums that are useful for at least one of its leading demanders. That is, each user $i$ broadcasts all $Y^i_{\mathcal{A}^i}$ for all subsets $\mathcal{A}^i$  that satisfy $\mathcal{A}^i\cap\mathcal{U}^i\neq\emptyset$, i.e. $X_i = \{Y^i_{\mathcal{A}^i}\}_{\mathcal{A}^i\cap\mathcal{U}^i\neq\emptyset}$. For each user $i \in [K]$, the size of the broadcasted codeword amounts to $\binom{K-1}{t}-\binom{K-1-N_{\textup{e}}(\boldsymbol{d}_{\backslash\{i\}})}{t}$ times the size of a sub-piece, summing which for all $i \in [K]$ results in the load stated in \eqref{eq:single}.

We now show that each user $k \in [K]$ is able to recover its desired sub-pieces. When $k$ is a leading demander of a user $i$, i.e., $k\in \mathcal{U}^i$, it can decode any sub-piece $W_{d_k, \mathcal{B}^k \cup \{i\},i}$, for any $\Bc^k \subseteq \mathcal{A}^i\backslash\{k\}$, $|\Bc^k| = t-1$ , from $Y^i_{\Bc^k \cup \{k\}}$ which is broadcasted from user $i$, by performing 
\begin{equation}\label{eq:decodeBroadcast}
W_{d_k, \mathcal{B}^k \cup \{i\},i} =  \left(\underset{x\in \mathcal{B}^k}{{\bigoplus}}W_{d_k, \{\mathcal{B}^k \cup \{i,k\}\}\backslash\{x\}, i}\right) \bigoplus Y^i_{\Bc^k\cup \{k\}} 
\end{equation}
as can be seen from \eqref{eq:Broadcasts}.

However, when $k \notin \mathcal{U}^i$, not all of the corresponding codewords
$Y^i_{\Bc^k \cup \{k\}}$ for its required sub-pieces $W_{d_k, \mathcal{B}^k\cup \{i\},i}$ are directly broadcasted from user $i$. User $k$ can still decode its desired sub-piece by generating the missing codewords based on its received codewords from user $i$ (see \cite{arXivVersion} for the proof).

In the following, we provide a short demonstration of the above presented ideas.
\paragraph*{An example}\label{para:ex}	
Let us consider the case when $N=2, K=4, M=1, t=KM/N=2$ and $\boldsymbol{d} = (1,2,1,1)$. Notice that $N_{\textup{e}}(\boldsymbol{d}_{\backslash\{2\}})=1$ and $N_{\textup{e}}(\boldsymbol{d}_{\backslash\{i\}})=2\; \text{for}\,\, i \in \{1,3,4\}.$ Each file is divided into $\binom{4}{2} = 6$ sub-files and users cache the following sub-files for each $i \in \{1,2\}$:
\begin{align*}
\Mc_1 = \:\:\:& \lbrace W_{i,\{1,2\}}, W_{i,\{1,3\}},\, W_{i,\{1,4\}} \rbrace,\\
\Mc_2 = \:\:\:& \lbrace W_{i,\{1,2\}}, W_{i,\{2,3\}},\, W_{i,\{2,4\}} \rbrace,\\
\Mc_3 = \:\:\:& \lbrace W_{i,\{1,3\}}, W_{i,\{2,3\}},\, W_{i,\{3,4\}} \rbrace,\\
\Mc_4 = \:\:\:& \lbrace W_{i,\{1,4\}}, W_{i,\{2,4\}},\, W_{i,\{3,4\}} \rbrace.
\end{align*}

After splitting the sub-files into $2$ equal length sub-pieces, users $1,3,4$ transmit the codewords 
$X_1 = \lbrace Y^1_{\{2,3\}}, Y^1_{\{2,4\}}, Y^1_{\{3,4\}} \rbrace,
X_3 = \lbrace Y^3_{\{1,2\}}, Y^3_{\{1,4\}}, Y^3_{\{2,4\}} \rbrace,
X_4 = \lbrace Y^4_{\{1,2\}}, Y^4_{\{1,3\}}, Y^4_{\{2,3\}} \rbrace,$
where $Y_{\mathcal{A}^i}^i$ is given by \eqref{eq:Broadcasts}.

Notice that for these users, there exists no subset $\Ac^i$ s.t. $\Ac^i \subseteq [K]\backslash\{i\}$, $|\Ac^i|=t=2$ which satisfies $\Uc^i \cap \Ac^i \neq \emptyset$. However, depending on the choice of $\Uc^2$, user 2 can find $\binom{K-1-N_{\textup{e}}(\boldsymbol{d}_{\backslash\{2\}})}{t}=1$  subset $\Ac^2$ with $\Uc^2 \cap \Ac^2 \neq \emptyset$. Such an $\Ac^2$ can be determined as $\{3,4\}, \{1,4\}, \{1,3\}$ for the cases of $\Uc^2= \{1\}$, $\Uc^2 = \{3\}$, $\Uc^2 = \{4\}$, respectively.

Picking user $1$ as its leading demander, i.e., $\mathcal{U}^2 = \{1\}$, user $2$ only transmits $X_2 = \lbrace Y^2_{\{1,3\}}, Y^2_{\{1,4\}} \rbrace$
sparing the codeword $Y^2_{\{3,4\}} = W_{1,\{2,3\},2} \Large{\oplus} W_{1,\{2,4\},2}$. As mentioned before, the choice of the leading demanders is arbitrary and any one of the $Y^2_{\{1,3\}},\, Y^2_{\{1,4\}},\, Y^2_{\{3,4\}}$ can be determined as the superfluous codeword. In fact, any one of these codewords can be attained by summing the other two, since $Y^2_{\{1,3\}} \Large{\oplus} Y^2_{\{1,4\}} \Large{\oplus} Y^2_{\{3,4\}} = 0$. 

From the broadcasted codewords, all users can decode all their missing sub-pieces by using the sub-pieces in their caches as side-information, by performing \eqref{eq:decodeBroadcast}.
As each sub-piece is composed of $F/t\binom{K}{t} = F/12$ bits and as $3 \times 3 + 1 \times 2 = 11$ codewords of such size are broadcasted, our scheme achieves a load of $11/12$, which could be directly calculated by \eqref{eq:single}.

\begin{remark}\label{rem:Ksharedlink}
	Notice that  the proposed scheme is in fact composed of $K$ shared-link models each with $N$ files of size $F' = F/K$ bits and $K' = K - 1$ users with caches of size $M' = \frac{N(t-1)}{(K-1)}$ units each. The corresponding parameter for each model is found to be $t' = \frac{K'M'}{N} = t-1$. Summing the loads of each $i \in [K]$ shared-link sub-systems ((3) in \cite{yu2018exact}) with parameters $F=F', K=K', M=M', t=t',  N_{\textup{e}}(\boldsymbol{d}) = N_{\textup{e}}(\boldsymbol{d}_{\backslash\{i\}})$,  yields \eqref{eq:single}.
\end{remark}

\begin{remark}\label{rem:JiConnection}
	When each user requests a distinct file ($N_{\textup{e}}(\boldsymbol{d}) = K$), our proposed scheme corresponds to the one presented in \cite{ji2016fundamental}. The potential improvement of our scheme when $N_{\textup{e}}(\boldsymbol{d}) < K$ hinges on identifying the possible linear dependencies among the codewords generated by a user.
\end{remark}

\section{Converse Bound under the Constraint of  One-Shot Delivery}\label{sec:Converse}
In this section we propose the converse bound under the constraint of one-shot delivery given in Theorem~\ref{teo}. Under the constraint of one-shot delivery, we can divide each sub-file $W_{i,\Vc}$ into sub-pieces. Recall that $W^{k,i}_{d_k,\Vc}$ represents the bits of $W_{d_k}$ decoded by user $k$ from $X_{i}$. 
Under the constraint of one-shot delivery, we can divide the D2D caching problem into $K$ shared-link models. In  the $i^{\textrm{th}}$ shared-link model where $i\in [K]$, user $i$  transmits $X_i$ such that each user $k\in [K]\setminus \{i\}$ can recover $W^{k,i}_{d_k,\Vc}$  for all $\Vc\subseteq ([K]\setminus \{k\})$ where $i\in \Vc$.

\subsection{Converse Bound for $R^*_{\textup{o}}(\mathbf{d},\boldsymbol{\mathcal{M}})$}
\label{sub:converse for R(d,M)}
Fix a demand vector $\mathbf{d}$ and a cache placement $\boldsymbol{\mathcal{M}}$.  We first focus on the shared-link model where user $i\in[K]$ broadcasts.

Consider a permutation of $[K]\setminus \{i\}$, denoted by $\mathbf{u}=(u_{1},u_{2},...,u_{K-1})$, where user $u_1$ is in position $1$ of $\mathbf{u}$, user $u_2$ is in position $2$ of $\mathbf{u}$, etc.
We define a function $f$ which maps the vectors $\mathbf{u}$ into another vector $f(\mathbf{u},\mathbf{d})$ based on the demand vector $\mathbf{d}$,
\begin{equation*}
f(\mathbf{u},\mathbf{d}):=\big(u_{j}: j\in [K-1] \textrm{ and } \{j^{\prime}\in[1:j-1]:d_{u_{j^{\prime}}}=d_{u_j}\}=\emptyset \big). 
\end{equation*}
In other words, to obtain $f(\mathbf{u},\mathbf{d})$, for each demanded file, from the vector $\mathbf{u}$ we remove  all the users  in $\mathbf{u}$ demanding this file except the user in the lowest position demanding this file.
For example, if $\mathbf{u}=(2,3,5,4)$, $\mathbf{d}=(1,2,2,3,3)$, we have $d_{u_1}=d_{u_2}=2$ and $d_{u_3}=d_{u_4}=3$, and thus $f(\mathbf{u},\mathbf{d})=(u_1, u_3)=(2,5)$. It can be seen that $f(\mathbf{u},\mathbf{d})$ contains $N_{\textup{e}}(\boldsymbol{d}_{\backslash\{k\}})$ elements.
Furthermore, we denote the $j^{\textrm{th}}$ element of $f(\mathbf{u},\mathbf{d})$ by $f_j(\mathbf{u},\mathbf{d})$. 
For the permutation $\mathbf{u}$, we can choose a set of sub-pieces, $\big(W^{f_j(\mathbf{u},\mathbf{d}),i}_{d_{f_j(\mathbf{u},\mathbf{d})},\Vc_{j}} : 
\Vc_{j}\subseteq[K]\backslash \{f_1(\mathbf{u},\mathbf{d}),\ldots,f_j(\mathbf{u},\mathbf{d})\}, \ i\in \Vc_j,
\ j\in[N_{\textup{e}}(\boldsymbol{d}_{\backslash\{i\}})]
\big)$. 
By a similar proof as~\cite[Lemma 1]{wan2016optimality}, we have the following lemma.
\begin{lemma}
	\label{lem:acyclic}
	For each permutation of $[K]\setminus \{i\}$, denoted by $\mathbf{u}=(u_{1},u_{2},...,u_{K-1})$, we have
	\begin{equation}
	H(X_i)
	\geq \label{eq:converse of Xk}\sum_{j\in[N_{\textup{e}}(\boldsymbol{d}_{\backslash\{i\}})]} \sum_{ \Vc_{j}\subseteq[K]\backslash \{f_1(\mathbf{u},\mathbf{d}),\ldots,f_j(\mathbf{u},\mathbf{d})\}: i\in \Vc_j}   |W^{f_j(\mathbf{u},\mathbf{d}),i}_{d_{f_i(\mathbf{u},\mathbf{d})},\Vc_{j}}|.		
	\end{equation}
\end{lemma}

Considering all the permutations of $[K]\setminus \{i\}$ and all $i\in [K]$, we sum the inequalities in form of~\eqref{eq:converse of Xk} to obtain,
\begin{equation}
(K-1)! \big(H(X_1)+\ldots+H(X_K) \big) \geq\label{eq:summing all Xk} \sum_{k\in[K]} \sum_{\Vc\subseteq [K]\setminus \{k\}} \sum_{i\in \Vc} a^{k,i}_{\Vc} |W^{k,i}_{d_k,\Vc}|,
\end{equation}
where $a^{k,i}_{\Vc}$  represents the coefficient of $|W^{k,i}_{d_k,\Vc}|$ in the sum. In \cite{arXivVersion}, we prove the following lemma.
\begin{lemma}
	\label{lem:same coeff}
	$a^{k,i_1}_{\Vc}=a^{k,i_2}_{\Vc}$, for each $i_1,i_2 \in \Vc$.
\end{lemma}
From Lemma~\ref{lem:same coeff}, we define $a^{k}_{\Vc}=\frac{ a^{k,i}_{\Vc}}{(K-1)! } $ for all $i\in \Vc$. Hence, from~\eqref{eq:summing all Xk} we have
\begin{align}
R^*_{\textup{o}}(\mathbf{d},\boldsymbol{\mathcal{M}})F\geq 
\big(H(X_1)&+\ldots+H(X_K) \big) \label{eq:sum to sub-file 1}\\
&  \geq \sum_{k\in[K]} \sum_{\Vc\subseteq [K]\setminus \{k\}} a^{k}_{\Vc} |W_{d_k,\Vc}|\label{eq:sum to sub-file}
\end{align} 
where in~\eqref{eq:sum to sub-file} we used 
\begin{align}
\sum_{i\in \Vc}|W^{k,i}_{d_k,\Vc}|\geq |W_{d_k,\Vc}|.\label{eq:sum to sub-file 3}\
\end{align}

\begin{remark}\label{rem:improvement}
	To derive the converse bound under the constraint of uncoded cache placement in~\cite{yu2018exact,wan2016caching},  the authors consider all the demands and all the permutations and sum the inequalities together. By the symmetry, it can be easily checked that in the summation expression, the coefficient of subfiles known by the same number of users is the same. However, in our problem, notice that~\eqref{eq:sum to sub-file 1} and~\eqref{eq:sum to sub-file 3} only hold for one demand. So each time we should consider one demand and let the coefficients of $H(X_k)$ where $k\in[K]$ be the same. Meanwhile, for each demand, we also should let the cofficients in Lemma~\ref{lem:same coeff} be the same.  
	However, for each demand, the $K$ shared-link models are not the symmetric. If we use the choice of the acyclic sets   in~\cite{yu2018exact,wan2016caching} for each of the K shared-link models,  we cannot ensure that for one demand, the coefficients   are  the symmetric. 
\end{remark}

\subsection{Converse Bound for $R^*_{\textup{ave, o}}$}
\label{sub:average converse}
We focus on a type of demands $\boldsymbol{s}$. For each demand vector $\mathbf{d} \in \Dc_{\boldsymbol{s}}$, we lower bound $R^*_{\textup{o}}(\mathbf{d},\boldsymbol{\mathcal{M}})$ as~\eqref{eq:sum to sub-file}. Considering all the demands in $\Dc_{\boldsymbol{s}}$, we then sum the inequalities in form of~\eqref{eq:sum to sub-file},
\begin{align}
\sum_{\mathbf{d} \in \Dc_{\boldsymbol{s}}} R^*_{\textup{o}}(\mathbf{d},\boldsymbol{\mathcal{M}})F\geq \sum_{q\in[N]} \sum_{\Vc\subseteq [K]} b_{q,\Vc} |W_{q,\Vc}|\label{eq:demand type}
\end{align}
where $b_{q,\Vc}$ represents the coefficient of $|W_{q,\Vc}|$. By the symmetry, it can be seen that $b_{q_1,\Vc_1}=b_{q_2,\Vc_2}$ if $|\Vc_1|=|\Vc_2|$. So we let $b_{t}:=b_{q,\Vc} $ for each $q\in [N]$ and $\Vc\subseteq [K]$ where $|\Vc|=t$. Hence, from~\eqref{eq:demand type} we get 
\begin{align}
|\Dc_{\boldsymbol{s}}|F \mathbb{E}_{\boldsymbol{d}\in \Dc_{\boldsymbol{s}}}[ R^*_{\textup{o}}(\boldsymbol{d},\boldsymbol{\mathcal{M}})]=\sum_{\mathbf{d} \in \Dc_{\boldsymbol{s}}} R^*_{\textup{o}}(\mathbf{d},\boldsymbol{\mathcal{M}})F\geq \sum_{t\in [0:K]} b_{t} x_t \label{eq:bt xt}
\end{align}
where we define $x_t:= \sum_{q\in [N]} \sum_{\Vc \subseteq [K]:|\Vc|=t} |W_{q,\Vc}|.$ The  value of $b_t$ is found as (see \cite{arXivVersion})
\begin{align}
b_t=\frac{|\Dc_{\boldsymbol{s}}| \Big( \sum_{i\in [K]} \binom{K-1}{t}-\binom{K-N_{\textup{e}}(\boldsymbol{d}_{\backslash\{i\}})-1}{t}\Big)}{tN\binom{K}{t}}.\label{eq:bt}
\end{align}
We take~\eqref{eq:bt} into~\eqref{eq:bt xt} to obtain
\begin{equation}
\mathbb{E}_{\boldsymbol{d}\in \Dc_{\boldsymbol{s}}}[ R^*_{\textup{o}}(\boldsymbol{d},\boldsymbol{\mathcal{M}})] \geq  \sum_{t\in [0:K]} \frac{ \binom{K-1}{t}- \frac{1}{K} \sum_{i\in [K]}\binom{K-N_{\textup{e}}(\boldsymbol{d}_{\backslash\{i\}})-1}{t} }{\binom{K-1}{t-1} NF}x_t
.\label{eq:take bt into xt 4}
\end{equation}
We also have the constraint of file size $\sum_{t\in [0:K]} x_t=NF,$ and the constraint of cache size $\sum_{t\in [1:K]} t x_t \leq KMF$. 

We let $r_{t,\boldsymbol{s}}:=\frac{ \binom{K-1}{t}- \frac{1}{K} \sum_{i\in [K]}\binom{K-N_{\textup{e}}(\boldsymbol{d}_{\backslash\{i\}})-1}{t} }{\binom{K-1}{t-1} NF}$. Similar to~\cite{yu2018exact}, we can lower bound~\eqref{eq:take bt into xt 4} using Jensen's inequality and the
monotonicity of $\textrm{Conv}(r_{t,\boldsymbol{s}})$,
\begin{align}
\mathbb{E}_{\boldsymbol{d}\in \Dc_{\boldsymbol{s}}}[ R^*_{\textup{o}}(\boldsymbol{d},\boldsymbol{\mathcal{M}})]\geq \textrm{Conv}(r_{t,\boldsymbol{s}}).
\end{align}
So we have
\begin{align}
\min_{\boldsymbol{\mathcal{M}}}\mathbb{E}_{\boldsymbol{d}\in \Dc_{\boldsymbol{s}}}[ R^*_{\textup{o}}(\boldsymbol{d},\boldsymbol{\mathcal{M}})]&\geq \min_{\boldsymbol{\mathcal{M}}} \textrm{Conv}(r_{t,\boldsymbol{s}})=\textrm{Conv}(r_{t,\boldsymbol{s}}).\label{eq:conv}
\end{align}
Considering all the demand types and from~\eqref{eq:conv}, we have
\begin{align}
R^*_{\textup{ave, o}}&\geq \mathbb{E}_{\boldsymbol{s}}\left[ \min_{\boldsymbol{\mathcal{M}}} \mathbb{E}_{\boldsymbol{d}\in \Dc_{\boldsymbol{s}}}[ R^*_{\textup{o}}(\boldsymbol{d},\boldsymbol{\mathcal{M}})] \right]\geq \mathbb{E}_{\boldsymbol{s}}[\textrm{Conv}(r_{t,\boldsymbol{s}})].\label{eq:consider ave}
\end{align}
Since $r_{t,\boldsymbol{s}}$ is convex, we can change the order of the expectation and the Conv in~\eqref{eq:consider ave}, to obtain Theorem~\ref{teo}.   Similarly, we can derive the converse bound on the worst-case load.

\section{Numerical Evaluations and Conclusions}	
We compare the load achieved by the presented one-shot scheme with the achievable load in \cite{ji2016fundamental} and with the minimum achievable load for the shared-link model \cite{yu2018exact}. We also provide the converse bounds in \cite{ji2016fundamental,sengupta2015beyond}.
\begin{figure}[t]
	\centering
	\includegraphics[width=0.82\textwidth]{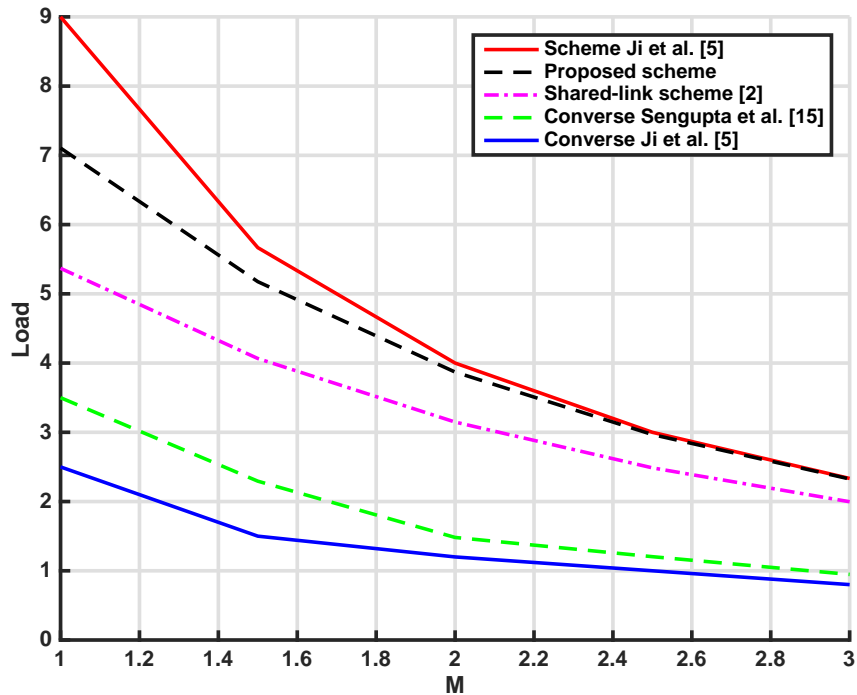}
	\caption{$N=10, K=20$, worst-case demand}
	\label{fig:N10K20w}
\end{figure}

In this work, we  characterized the load-memory
trade-off for cache-aided D2D networks under the constraints of uncoded placement and one-shot delivery. We presented a caching scheme and
proved its exact optimality in terms of both average and peak loads. The presented scheme is optimal within a factor of $2$, when the constraint of one-shot delivery is removed \cite{arXivVersion}.

\bibliographystyle{IEEEtran}
\bibliography{pub}

				\end{document}